\documentstyle[12pt,extras,epsfig,citews]{article}

\newcommand{\bcn}{\begin{center}}
\newcommand{\beq}{\begin{equation}}
\newcommand{\beqn}{\begin{eqnarray}}
\newcommand{\ecn}{\end{center}}
\newcommand{\eeq}{\end{equation}}
\newcommand{\eeqn}{\end{eqnarray}}
\newcommand{\case}[2]{\mbox{\small $\displaystyle \frac{#1}{#2}$}}
\newcommand{\dcsb}{\mbox{D$\chi$SB}}

\newcommand{\Eq}[1]{Eq.~(\ref{#1})}

\newcommand{\qbq}{\mbox{$\langle\overline{q}q\rangle$}}

\newcommand{\sect}[2]{\vspace*{6mm}\hspace*{-\parindent}{\bf #1.}~{\bf
#2}\vspace*{4mm}}
\newcommand{\subsect}[2]{\vspace*{5mm}\hspace*{-\parindent}{\bf #1}~{\it
#2}\vspace*{3mm}}
 \def\lsim{\mathrel{\rlap{\lower4pt\hbox{\hskip1pt$\sim$}}
    \raise1pt\hbox{$<$}}}         
\begin{document}
\rightline{ANL Preprint Number: PHY-7853-TH-94}
\bcn
{\bf \footnote[2]{Summary of a presentation at {\it The Workshop on Quantum
Infrared Physics}, The American University of Paris, 6-10 June, 1994.}CAN AN
INFRARED-VANISHING GLUON PROPAGATOR CONFINE
QUARKS?
}\vspace*{1cm}

{\bf CRAIG D. ROBERTS}\\

{\it Physics Division, Bldg. 203, Argonne National Laboratory \\
        Argonne, IL 60439-4843, USA}\vspace*{1cm}

{\bf ABSTRACT}
\ecn
\hspace*{3pc}
\parbox{30pc}{\small
It is shown that the solution of the quark Dyson-Schwinger equation in QCD
obtained with a gluon propagator of the form $D(q) \sim q^2/[q^4+b^4]$ and a
quark-gluon vertex that is free of kinematic singularities does not describe
a confined particle and that there is always a value of $b^2 = b^2_c$ such
that chiral symmetry is not dynamically broken for $b^2>b_c^2$.}

\sect{1}{Introduction}

Dynamical Chiral Symmetry Breaking (\dcsb) and confinement are two crucial
features of quantum chromodynamics (QCD).  They are responsible for the
nature of the hadronic spectrum; \dcsb\ ensuring the absence of low mass
scalar partners of the pion and confinement ensuring the absence of free
quarks, for example.  A natural method for studying both \dcsb\ and
confinement in QCD is the complex of Dyson-Schwinger Equations
(DSEs).\cite{DSErev}

One goal of DSE studies is to develop this nonperturbative approach to the
point where it is as firmly founded as lattice QCD and calculationally
competitive.  Although more needs to be done in order to achieve this goal
there has been a good deal of progress, especially in the study of Abelian
gauge theories where direct and meaningful comparisons can be made, and
agreement obtained, between the results of lattice and DSE
studies.\cite{QEDCJB}

Herein a recent study\cite{HRW94} of the fermion DSE in which the gluon
propagator vanishes at $q^2=0$, aimed at determining whether such a gluon
propagator can support \dcsb\ and/or generate a confining quark propagator,
is described.  There has been renewed interest in such a form of the gluon
propagator, which was argued in Ref.~\citenum{VG79} to be associated with the
elimination of Gribov copies, because of the recent work of H\"abel {\it
et. al.}\cite{H90b} and Zwanziger.\cite{Zw91}

In Sec.~2 the DSE for the quark propagator is described in detail.  There is
also a discussion of what is known about the dressed gluon propagator and
quark-gluon vertex in QCD.  The analysis of \dcsb\ is reported in Sec.~3 and
quark confinement is discussed in Sec.~4.  The results are summarised in
Sec.~5.

\sect{2}{Dyson-Schwinger Equation for the Fermion Self Energy}

\begin{figure}[htb] 
  \centering{\ \epsfig{figure=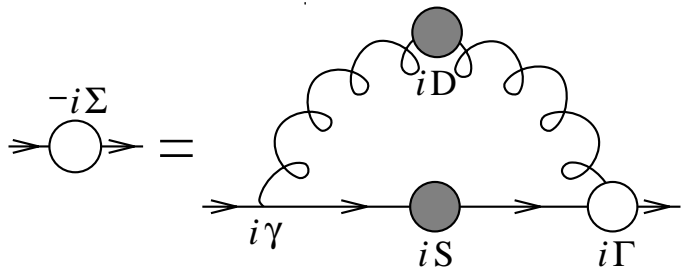,height=3.5cm} }
\parbox{130mm}{\caption{The Dyson-Schwinger equation for
the quark self-energy.
\label{quark_dse_fig} }}
\end{figure}

In Minkowski space, with metric \mbox{$g_{\mu\nu} = \;{\rm diag}
(1,-1,-1,-1)$} and in a general covariant gauge, the inverse of the dressed
quark propagator can be written as
\begin{equation}
 S^{-1}(p) = \not\!{p} - m - \Sigma(p)
	   \equiv Z^{-1}(p^2) \left( \not\!{p} -M(p^2) \right)
           \equiv A(p^2) \not\!{p} - B(p^2)\/,
\end{equation}
with: $m$\ the renormalised, explicit chiral symmetry breaking mass (if
present); $\Sigma(p)$\ the self-energy; $M(p^2) = B(p^2)/A(p^2)$\ the
dynamical quark mass function; and $Z(p^2) = A^{-1}(p^2)$ the
momentum-dependent renormalisation of the quark wavefunction.  The
unrenormalised DSE for the inverse propagator is
\begin{equation}
 S^{-1}(p) = \not\!{p} - m^{\rm bare}
     - i \case{4}{3} g^2 \int \frac{d^4k}{(2\pi)^4} \gamma^\mu
         S(k) \Gamma^\nu (k,p) D_{\mu \nu}((p-k)^2)\/,     \label{fullDSE}
\end{equation}
where $D_{\mu\nu}(q^2)$\ is the dressed gluon propagator and $\Gamma^\nu$ is
the proper quark-gluon vertex, which is illustrated in
Fig.~\ref{quark_dse_fig}.

The renormalised, massless ($m=0$) DSE can be written as
\begin{equation}
 \Sigma_R(p)=   \left(1-Z_S\right)\not\!{p} +
        i Z_\Gamma \case{4}{3}  g^2 \int^{\Lambda}
         \frac{d^4k}{(2\pi)^4} \gamma^\mu
         S_R(k) \Gamma_R^\nu (k,p) D^R_{\mu \nu}((p-k)^2)\/,
\label{ModRDSE}
\end{equation}
where $Z_S$ and $Z_\Gamma$ are quark-propagator and quark-gluon-vertex
renormalisation constants, respectively, which depend on the renormalisation
scale, $\mu$, and ultraviolet cutoff, $\Lambda$.  Hereafter we suppress the
label $R$.

The solution of this equation provides information about \dcsb\ .  The quark
condensate, $\qbq \propto \mbox{tr$[S(x=0)]$}$, is a chiral symmetry order
parameter.  If there is a solution of \Eq{ModRDSE} with $B\neq 0$ then the
quark has generated a mass via interaction with its own gluon field and the
chiral symmetry is therefore dynamically broken.  The solution also provides
information about quark confinement, as discussed in Sec.~4.

\subsect{2.1a}{Gluon Propagator}

In a general covariant gauge the dressed gluon propagator, which is diagonal
in colour space, can be written:
\begin{equation}
\label{Gprop}
D^{\mu\nu}(q^2) = \left[ \left(g^{\mu\nu}-\frac{q^\mu q^\nu}{q^2}\right)
  \frac{1}{1-\Pi(q^2)} + \xi \frac{q^\mu q^\nu}{q^2} \right]
\frac{1}{q^2}\/,
\end{equation}
where $\Pi(q^2)$\ is the gluon vacuum polarisation and $\xi$\ is the gauge
parameter.  In covariant gauges the longitudinal piece of this propagator is
not modified by interactions, which follows from the Slavnov-Taylor
identities in QCD.

\begin{figure}[htb] 
 \centering{\ \epsfig{figure=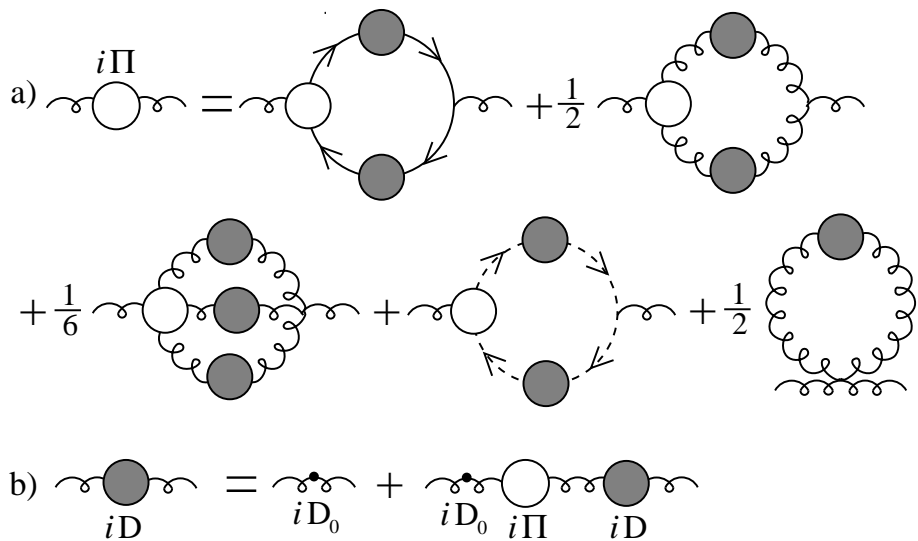,height=7.0cm} } \parbox{130mm}{
\caption{ The Dyson-Schwinger equation for the gluon propagator.
[Here and below the broken line represents the
propagator for the ghost field.]
\label{gluon_dse_fig}  }}
\end{figure}

The Dyson-Schwinger equation for the gluon propagator is given
diagrammatically in Fig.~\ref{gluon_dse_fig}.
The symmetrisation factors of 1/2 and 1/6 arise from the usual Feynman rules,
which also require a negative sign [unshown] to be included for every fermion
and ghost loop.  This equation has been studied
extensively.\cite{H90b,SM79,UBG80,BBZBP}  There have also been attempts to
determine the gluon propagator from numerical simulations of
lattice-QCD.\cite{MO,BPS94}

\begin{figure}[htb] 
\centering{\
\epsfig{figure=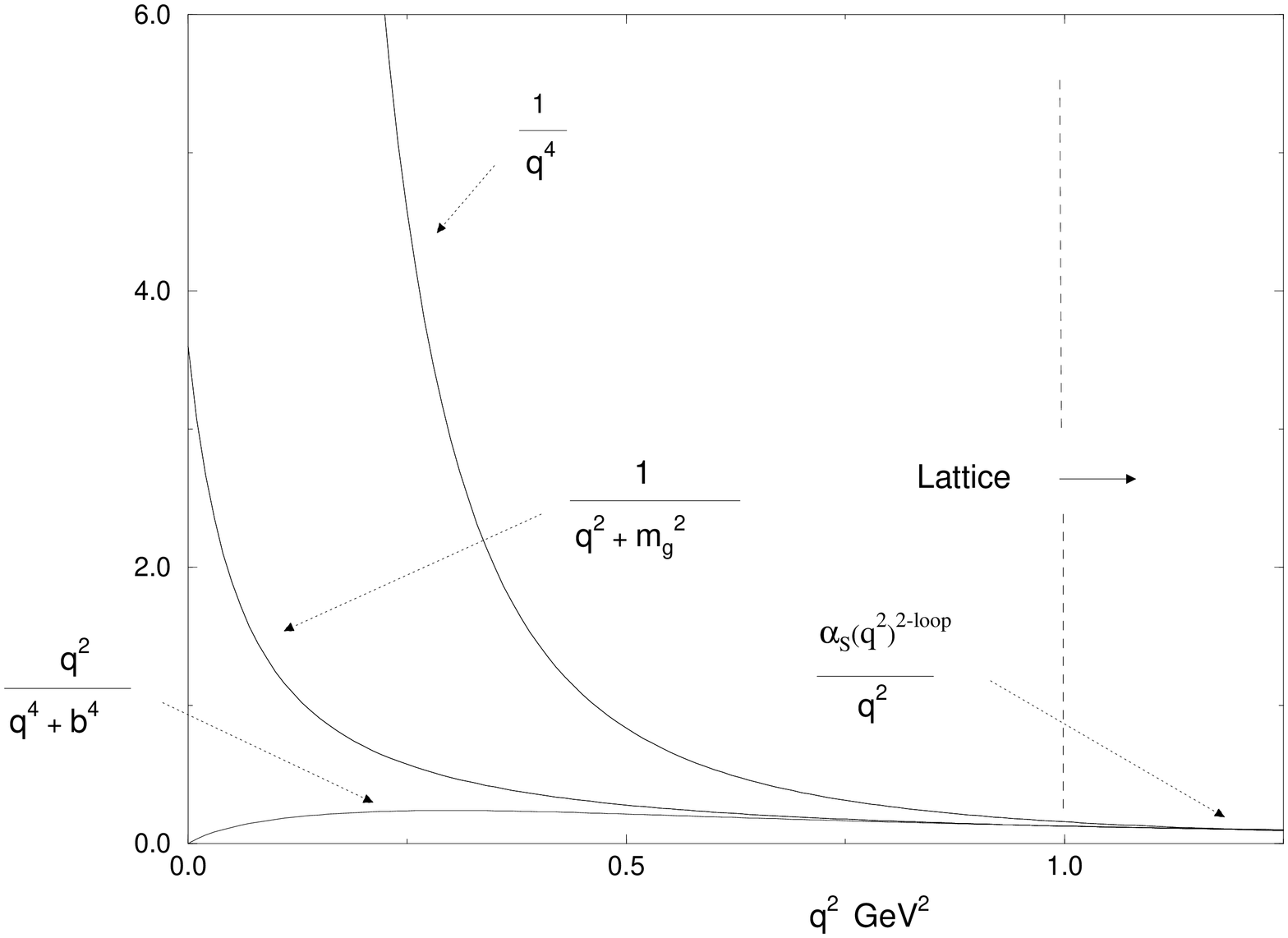,height=13.5cm,angle= -90} }
\parbox{130mm}{
\caption{Results of studies of the gluon propagator in QCD.  Typical values
of $m_g = 0.5$~GeV and $b= 0.7$~GeV have been used.
\label{plot_D} } }
\end{figure}

The results of the DSE and lattice studies are summarised in Sec.~5.1 of
Ref.~\citenum{DSErev} and are represented in Fig.~\ref{plot_D}.
This figure illustrates that for spacelike-$q^2> 1$~GeV$^2$ the gluon
propagator is given by the two-loop, QCD renormalisation group result, with
the next order correction being $<$10\%.  For spacelike-$q^2< 1$~GeV$^2$,
however, the form of the propagator is not known. The DSE studies of
Refs.~\citenum{SM79,UBG80,BBZBP} suggest a regularised infrared singularity,
represented by $1/q^4$ in the figure. That of Ref.~\citenum{H90b}, which
differs mainly in that the Ansatz used for the triple-gluon vertex has
kinematic singularities, suggests an infrared vanishing form, characterised
by $q^2/(q^4+b^4)$, which has also been argued\cite{VG79,Zw91} to be the form
necessary to completely eliminate Gribov copies.

The lattice Landau-gauge studies of Ref.~\citenum{MO} favour the massive
vector boson form, $1/(q^2+m_g^2)$, which is broadly consistent with the
improved studies of Ref.~\citenum{BPS94}.  On $16^3\times 40$ and $24^3\times
40$ lattices at $\beta=6.0$ these numerical results allowed a fit of the form
$q^2/(q^4+b^4)$, with $b\sim 340$~MeV, but a fit using a standard massive
particle propagator could not be ruled out. On a lattice of dimension
$16^3\times 24$ at $\beta=5.7$ it was found that the gluon propagator was
best fit by a standard massive vector boson propagator with mass $\sim
600$~MeV.  There is a problem with these studies, however, which is indicated
by the dashed vertical line at the right of Fig.~\ref{plot_D}.  With present
technology, the domain of spacelike-$q^2< 1$~GeV$^2$ is actually inaccessible
in lattice studies and, since all forms of the propagator are very nearly the
same outside this domain, it is clear that these results are both
qualitatively and quantitatively unreliable.

This brief discussion indicates that at present one can only say that
\begin{equation}
D(q^2) \equiv \frac{1}{q^2\,[1-\Pi(q^2)]} = \frac{q^2}{q^4 + b^4}
\label{resb}
\end{equation}
is not implausible in QCD, at least at small $q^2$.  For this reason it is of
interest to determine whether such a form of $D(q^2)$ can lead to \dcsb\ and
a confining quark propagator, which will provide further insight into the
validity of this form of gluon propagator.  In keeping with the ultraviolet
behaviour illustrated in Fig.~\ref{plot_D}, the results obtained with the
``ultraviolet-improved'' form:
\beq
\frac{g^2}{4\pi}\,D(q^2) = \alpha(\tau;q^2)\,\frac{q^2}{q^4 + b^4}~,
\label{Dq}
\eeq
with
\mbox{$
\alpha(\tau;q^2) = (d\pi)/\left(\ln\left[\tau +
q^2/\Lambda_{QCD}^2\right]\right)$}, \mbox{$d=12/[33-2 N_f]$}, $N_f = 4$,
where $\tau\geq 1$ is an infrared regularising parameter, are described
herein.

\subsect{2.1b}{Gluon Condensate}

The gluon condensate can be calculated from the nonperturbative part of the
gluon propagator.\cite{RWK92}  Using \Eq{Dq} one obtains
\beq
\langle \alpha_S GG\rangle_{\mu}
= -\,\frac{3b^4}{\pi^2}\,
\ln\left(\ln\left[\frac{\mu^2}{\Lambda_{QCD}^2}\right]\right)~.
\eeq
For $b=340$~MeV, $\Lambda_{QCD} = 200 MeV$ and $\mu=1$~GeV this yields
$\langle \alpha_S GG\rangle_{\mu} = -0.0047$~GeV$^4$, which should be
compared with the value inferred from QCD sum rules: $\langle
\alpha_S GG\rangle_{\mu} \sim 0.04$~GeV$^4$.  The ``wrong sign'' is
due to the fact that \Eq{Dq} is weaker than perturbative gluon exchange for
all spacelike-$q^2$ and is an harbinger of the results to follow.

\subsect{2.2}{Quark-Gluon Vertex}

The quark-gluon vertex satisfies its own DSE but hitherto there have been no
studies of this equation in QCD.  Making use of the ``Abelian
approximation'',\cite{UBG80} the Slavnov-Taylor identity for this vertex
reduces to the Ward-Takahashi identity familiar from QED:
\begin{equation}
\label{WI}
k_\mu\Gamma^\mu(p,q)= S^{-1}(p) - S^{-1}(q)~,
\end{equation}
$k= (p-q)$, and the quark-gluon vertex can be written in the general
form\cite{BC80}
\beq
\label{GV}
\Gamma_\mu(p,q) = \Gamma_\mu^{\rm BC}(p,q)
        + \sum_{i=1}^8\, T_\mu^i(p,q)\,g^i(p^2,p\cdot q,q^2)~,
\eeq
where the eight tensors, $T^i_\mu$, are transverse, $(p-q)^\mu
T_\mu^i(p,q)=0$ and
\beqn
\Gamma_\mu^{\rm BC}(p,q) & = & \Sigma_A(p,q)\,\gamma_\mu
        + (p+q)_{\mu}\left\{ \Delta_A(p,q)\,\case{1}{2}\,
                \left[ \gamma\cdot p + \gamma\cdot q\right]
- i\Delta_B(p,q)\right\}
\eeqn
with $\Sigma_F(p,q) \equiv[F(p^2) + F(q^2)]/2$ and $\Delta_F(p,q) \equiv
[F(p^2) - F(q^2)]/[p^2 - q^2]$, for $F=A$ or $B$.  In this approximation QED
studies~\cite{DMR94} can be used to place constraints on the functions
$g^i$. Taking these constraints into account leads to the Ansatz:\cite{CP}
$T^i_\mu = 0$, $i\neq 6$, and
\beqn
\label{CPV}
T_\mu^6(p,q)\,g^6(p^2,p\cdot q,q^2) & =&
\frac{\gamma^\mu(k^2-p^2) - (k+p)^\mu (\not\!{k}-\not\!{p})}{2d(k,p)}
      \left[ A(k^2)-A(p^2) \right]\/,             \label{CPGamT} \\
  {\rm with}~d(k,p) & =& \frac{1}{(k^2+p^2)}\left( (k^2-p^2)^2
	   + \left[ \frac{B^2(k^2)}{A^2(k^2)}+\frac{B^2(p^2)}{A^2(p^2)}
	      \right]^2 \right).
\eeqn

This Ansatz, \Eq{GV} with \Eq{CPGamT} and $T^i_\mu = 0$ for $i\neq 6$,
satisfies the Ward-Takahashi Identity, is free of kinematic singularities
(i.e., has a well defined limit as ($p\rightarrow q$), reduces to the bare
vertex in the free field limit in the manner prescribed by perturbation
theory, transforms correctly under charge conjugation and Lorentz
transformations and preserves multiplicative renormalisability in the quark
DSE.  (Of these properties the vertex $\Gamma_\mu^{\rm BC}$ satisfies all but
the last and, for the most part, the calculations described herein were
performed with the Ansatz $\Gamma_\mu =
\Gamma_\mu^{\rm BC}$.)

It should be noted that the absence of kinematic singularities is an
important and physically reasonable constraint.  To understand this one can
simply consider an analogy with PCAC.  In the case of the axial-current
vertex, $\Gamma_\mu^5$, there is a kinematic singularity in the chiral limit,
which is identified with the massless pion excitation.  A kinematic
singularity in the quark-gluon vertex would therefore entail the existence of
an hitherto unknown, massless excitation in QCD.

\subsect{2.3}{Quark DSE in Euclidean Space}

In the Abelian approximation one has \mbox{$Z_S=1=Z_\Gamma$} at one-loop in
Landau gauge.  Using this result here considerably simplifies \Eq{ModRDSE},
which can be written in Euclidean space, with metric
\mbox{$\delta_{\mu\nu}={\rm diag}(1,1,1,1)$} and $\gamma_\mu$ hermitian,
\begin{equation}
 \Sigma(p)= \case{4}{3} g^2 \int^{\Lambda} \frac{d^4k}{(2\pi)^4} \gamma_\mu
S(k) \Gamma_\nu (k,p) D_{\mu \nu}((p-k)^2) \label{EDSE}
\end{equation}
where
\begin{equation}
S^{-1}(p) = i\gamma\cdot p + \Sigma(p) = i\gamma\cdot p A(p^2) + B(p^2)
\end{equation}
and all the other elements in this equation are taken to be specified by the
expressions given above evaluated at Euclidean (spacelike) values of their
arguments.  The important, subtle considerations associated with the
transformation between Minkowski and Euclidean space are discussed in
Sec.~2.3 of Ref.~\citenum{DSErev}.

Equation~(\ref{EDSE}) is actually a pair of coupled, nonlinear integral
equations for $A(p^2)$ and $B(p^2)$.  With $\Gamma_\mu = \Gamma_\mu^{\rm BC}$
one obtains
\beqn
\label{Beqn}
\lefteqn{B(p^2)  =  \frac{16\pi}{3} \int^\Lambda \frac{d^4k}{(2\pi)^4}
        \alpha(\tau;(p-k)^2)\frac{(p-k)^2}{(p-k)^4 + b^4}\,
        \frac{1}{A^2(k^2)k^2+B^2(k^2)} \times} \\
  & &
        \left\{3B(k^2)\frac{A(k^2)+A(p^2)}{2}
 + \left[B(k^2)\Delta A(k^2,p^2)
-A(k^2)\Delta B(k^2,p^2)\right]h(p,k)\right\}\/,
\nonumber
\eeqn
with \mbox{$h(p,k)=2\left[k^2p^2-(k\cdot p)^2\right]/(p-k)^2$}, and a similar
but more complicated equation for $A(p^2)$, while with $\Gamma_\mu =
(\Gamma_\mu^{\rm BC} + T_\mu^6\,g^6$), from \Eq{CPV}, one has
\beqn
\lefteqn{B(p^2)  =  {\rm RHS~of~(\ref{Beqn})} }\label{BT} \\
 && \!\!\!\!\!\!\!\!\!\!
+ \frac{16\pi}{3}\int^\Lambda \frac{d^4k}{(2\pi)^4}
        \alpha(\tau;(p-k)^2)\frac{(p-k)^2}{(p-k)^4 + b^4}\,
        \frac{B(k^2)\Delta A(k^2,p^2) }{A^2(k^2)k^2+B^2(k^2)}
    \frac{(k^2-p^2)}{2d(k,p)}
    3(k^2-p^2)~,\nonumber
\end{eqnarray}
with, again, a similar but more complicated equation for $A(p^2)$.

The results discussed below were obtained from an iterative, numerical
solution of these equations on a logarithmic grid of
\mbox{$x= p^2/\Lambda_{\rm QCD}^2$} and
\mbox{$y=k^2/\Lambda_{\rm QCD}^2$} points.   The solutions  were independent
of the seed-solution and grid choice and also of the UV cutoff, which was
\mbox{$\Lambda^2 = 5\times 10^8\;\Lambda_{\rm QCD}^2$}.

\sect{3}{Dynamical Chiral Symmetry Breaking?}

The gauge-invariant quark condensate is an order parameter for \dcsb\ and is
obtained from the trace of the quark propagator:
\begin{equation}
\langle \overline{q}q\rangle_\mu = -\frac{3}{4\pi^2}
\ln\left(\frac{\mu^2}{\Lambda_{\rm QCD}^2}\right)^{d}\,
\lim_{\Lambda^2\rightarrow \infty} \left(
\ln\left(\frac{\Lambda^2}{\Lambda_{\rm QCD}^2}\right)^{-d}\,
\int_0^{\Lambda^2}
ds\, s\,\frac{B(s)}{s A(s)^2 + B(s)^2}\right)~,
\label{cndst}
\end{equation}
where $\mu$ is the renormalisation point for the condensate, which is usually
fixed at \mbox{$1$ GeV}.  This is the parameter that is used to study
\dcsb\  in lattice QCD.  A nonzero value signals \dcsb.

\begin{figure}[htb] 
  \centering{\ \epsfig{figure=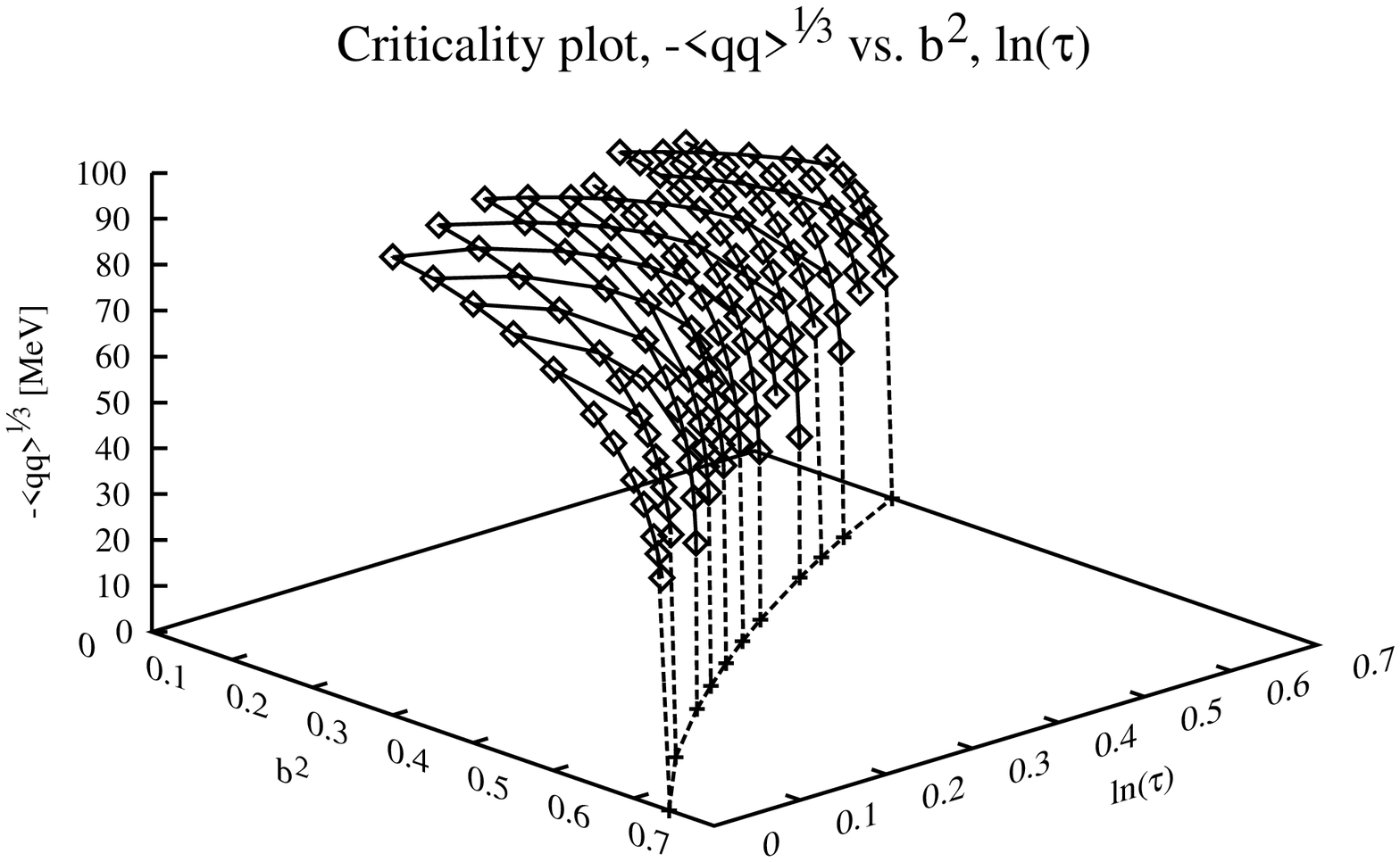,height=10.0cm} }
\parbox{130mm}{\caption{ Criticality plot for
  $(-\langle \bar{q}q \rangle_\mu)^{\frac{1}{3}}$\ as a function of $\ln
\tau$\ and $b^2$\/.  The condensate, $(-\langle \bar{q}q
\rangle_\mu)^{\frac{1}{3}}$, is in units of MeV, scaled to $\mu^2 = 1 {\rm
GeV}$\/, and $b^2$\ is in units $\Lambda_{QCD}^2$\/ [ $b^2=0.49 \Rightarrow
b\sim 140$ MeV]; the gluon regulator $\tau$\ is dimensionless.
\label{dse_hrw_fig1}  }}
\end{figure}

In Fig.~\ref{dse_hrw_fig1} the condensate obtained from the numerical
solution of \Eq{Beqn} and the associated equation for $A(p^2)$ for values of
$\ln\tau$ in the domain \mbox{$[0.0,0.7]$} and $b^2$ in
\mbox{$[0.1,1.0]$} is plotted. This figure shows regions of unbroken and
dynamically broken chiral symmetry.  Notably there is no \dcsb\ for the value
of $b^2 \sim 3~\Lambda_{\rm QCD}^2$ inferred from lattice simulations.

The numerical results suggest that $\langle \bar q q \rangle_{\mu}$ rises
continuously from the transition boundary and hence that the transition is
second order.  Assuming therefore that, for a given value of $\ln\tau$, the
order parameter behaves as
\mbox{$
      \langle\overline{q}q\rangle_\mu(b^2) \approx C
\left(1-b^2/b^2_c\right)^{\beta}$}
one obtains $\beta = 0.572$ with $\sigma_\beta = 0.020$.  Including the
$T_\mu^6$ term in the vertex only leads to a small quantitative change in the
results.  For example, with $\ln\tau=0.6$ one finds $\beta_{T^6}=0.579$,
$\sigma_{\beta_{T^6}} = 0.015$.  This suggests that for any vertex that is
free of kinematic singularities there is a critical value of $b^2=b_c^2$ such
that there is no \dcsb\ for $b^2>b_c^2$.

\sect{4}{Quark Confinement?}

In order to determine whether the quark propagator obtained as a solution to
\Eq{EDSE} with an infrared-vanishing gluon propagator can represent a
confined particle we follow Ref.~\citenum{HRM92} and adapt a method commonly
used in lattice QCD to estimate bound state masses.  Writing
\mbox{$ \sigma_S(s) = B(s)/[s A(s)^2 + B(s)^2] $}, $s=p^2$,  defining
\beq
\Delta_S(T) = \int_{-\infty}^\infty\frac{dy}{2\pi}\,\sigma_S(y^2)\,
        {\rm e}^{i\,y\,T}
\eeq
and, for notational convenience,
\mbox{$E(T) = -  \ln\left[\Delta_S(T)\right]$}, it follows from the axioms of
field theory that if there is a {\it stable asymptotic state} with the
quantum number of the quark then
\begin{equation}
\label{ConfT}
\lim_{T\rightarrow\infty}
 \frac{dE(T)}{dT} \, = m~;
\end{equation}
where $m \geq 0$ is the mass of this excitation; i.e, this limit yields the
dynamically generated quark mass.

As a simple example one can consider the Nambu-Jona-Lasinio model\cite{NJL}
in which the dressed quark propagator is
\mbox{$S(p)=1/[i\gamma\cdot p + M]$} and hence
\mbox{$\sigma_S(s) = M/[s + M^2]$}.  In this case
\mbox{$\Delta_S(T) = \exp(-\,M\,T)\,/2$},  which, from \Eq{ConfT}, yields
$m=M$, as one would expect.

\begin{figure}[htb] 
 \centering{\
  \epsfig{figure=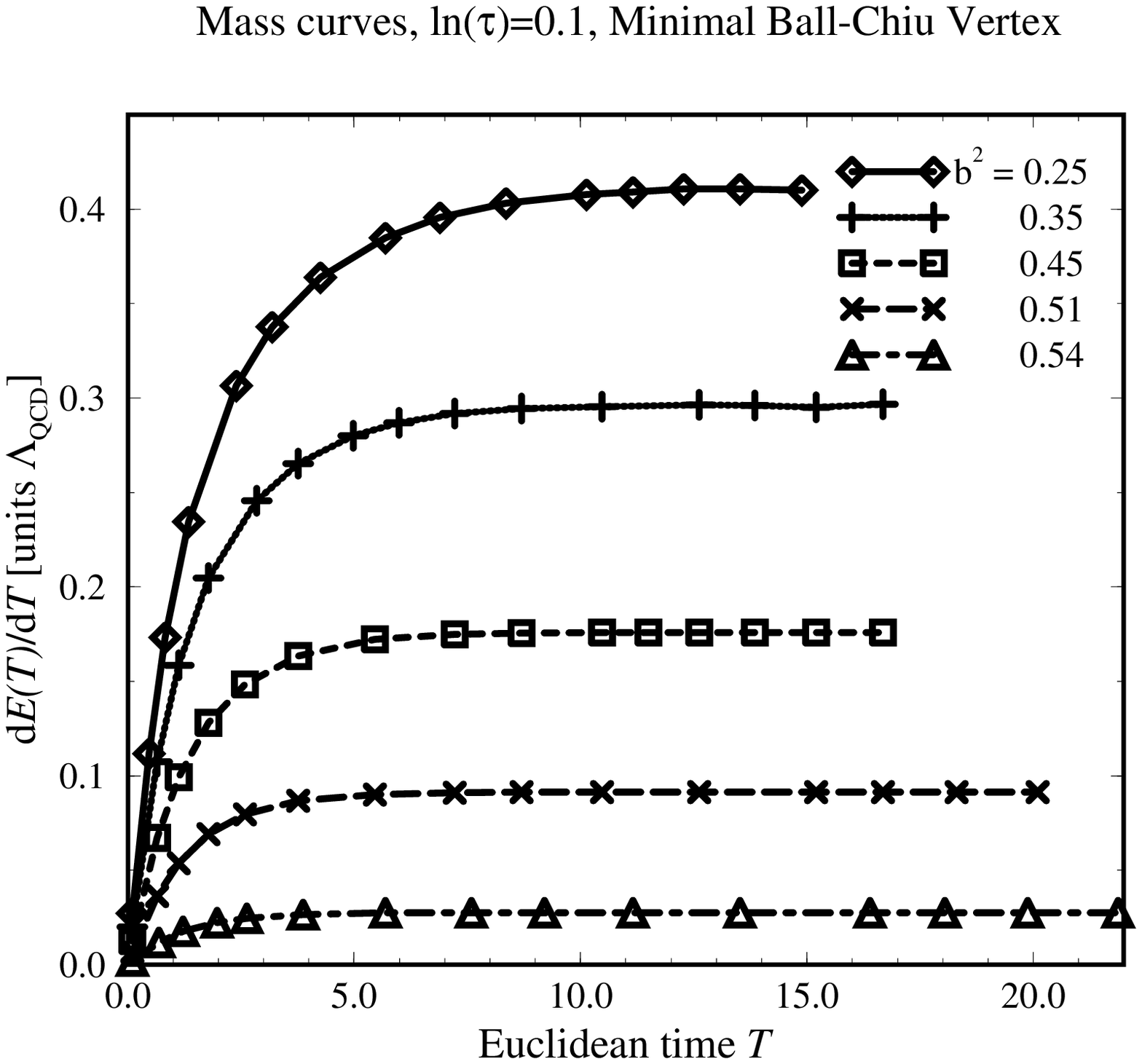,height=8.5cm} }
  \parbox{130mm}{
\caption{ Dressed-quark-mass curves for the family of propagators with
the minimal Ball-Chiu vertex and $\ln\tau=0.1$. The masses are in units
of $\Lambda_{\rm QCD}.$
\label{dse_hrw_fig4} }}
\end{figure}

This confinement test was applied to the numerical solutions in the following
cases: 1) The propagators obtained with \mbox{$\ln\tau=0.1$} and $b^2$ in the
range \mbox{$[0.1,1.0]$}; 2) The propagator obtained with \mbox{$\ln\tau =
0$} and \mbox{$b^2 = 0.35$}, which yields the largest value of
\mbox{$(-\langle\overline{q}q\rangle_\mu)$} on the \mbox{$(b^2,\ln\tau)$}
domain considered; 3) Two propagators obtained with
\mbox{$(b^2,\ln\tau) = (0.1,0.6)$} - one using $\Gamma_\mu=\Gamma_\mu^{\rm
BC}$ and the other using $\Gamma_\mu = (\Gamma_\mu^{\rm BC} + T_\mu^6\,g^6$).
Plots of \mbox{$E'(T)$} for the family of
propagators obtained with \mbox{$\ln\tau = 0.1$} are presented in
Fig.~\ref{dse_hrw_fig4}.
(It should be noted that a quark propagator with complex conjugate
poles\cite{H90b} leads to a form of $\Delta_S(T)$ with zeros and hence to
$E'(T)$ with zeros and poles; a strong signal of which there is no sign in
Fig.~\ref{dse_hrw_fig4}.)  Since the behaviour of all the other solutions was
qualitatively the same as that described by the results presented in
Fig.~\ref{dse_hrw_fig4} one concludes that, independent of $b$, $D(q)\sim
q^2/[q^4+b^4]$ does not yield a confining quark propagator.

\sect{5}{Summary}

A study of the quark Dyson-Schwinger equation in QCD using a gluon propagator
that vanishes as \mbox{$q^2\rightarrow 0$}, \Eq{Dq}, and a dressed
quark-gluon vertex, $\Gamma_\mu(p,q)$, that has no kinematic singularities;
i.e., has a well defined limit as $p\rightarrow q$, is described.  The
results indicate that such a gluon propagator can only support \dcsb\  for
values of $\ln\tau$ and $b^2$ less than certain critical values and does not
confine quarks.  The results are qualitatively independent of the model
dependent elements in this study.  One is therefore lead to conclude that the
dressed gluon propagator in QCD does not vanish in the infrared.

\sect{Acknowledgments}

I would like to thank and congratulate the organisers, Berndt M\"{u}ller and
Herb Fried, for bringing about this timely and stimulating meeting, and Dan
Zwanziger for giving me the opportunity to make this presentation in his
session.  This work was supported by the Department of Energy, Nuclear
Physics Division under contract number W-31-109-ENG-38.



\end{document}